\documentclass[11pt,letterpaper]{article}
\usepackage{fullpage}
\usepackage{authblk}
\usepackage[margin=1in,headsep=\baselineskip, headheight=17pt]{geometry}
\usepackage{setspace}
\usepackage{wrapfig}
\usepackage{graphicx}
\usepackage{cite}
\usepackage{verbatim} 
\usepackage{gensymb}
\usepackage{booktabs}
\usepackage{fancyhdr}
\usepackage{amsmath}
\usepackage{amsfonts}
\usepackage[font=footnotesize,labelfont=bf]{caption}
\usepackage{subcaption}
\usepackage{floatrow}
\usepackage{pdfpages}
\usepackage{amssymb,amsmath,amsthm}
\usepackage{bm}
\usepackage[english]{babel}
\usepackage{multirow}
\usepackage{enumitem}
\usepackage{float}
\usepackage{miller}
\usepackage{placeins}
\usepackage{blindtext}
\usepackage{physics}
\usepackage[table,xcdraw]{xcolor}
\usepackage[%  
colorlinks=true,
linkcolor=blue,
citecolor=blue,
urlcolor=blue
]{hyperref}
\usepackage{titlesec}
\usepackage{parskip}
\usepackage{graphicx}
\usepackage{fontawesome}
\usepackage{tabularx}
\newcolumntype{b}{X}
\newcolumntype{s}{>{\hsize=.6\hsize}X}

\title{\vspace{-3em}\bfseries Dynamic Mechanical Response of Spinodal Architectures Across Length and Time Scales}

\author[1]{Vatsa Gandhi\thanks{Corresponding author: vatsa@ucla.edu}$^{\dagger}$}
\author[2]{Rishi Kommalapati$^{\dagger}$}
\author[2]{Carlos M. Portela}
\author[3]{Vikram Deshpande}

\affil[1]{Mechanical and Aerospace Engineering Department, University of California, Los Angeles, Los Angeles, CA 90095, USA}
\affil[2]{Department of Mechanical Engineering, Massachusetts Institute of Technology, Cambridge, 02139, MA, USA}
\affil[3]{Department of Engineering, University of Cambridge, Cambridge CB2 1PZ, UK}

\date{}

\titleformat{\section}[hang]{\large\bfseries}{\thesection \hspace{0.5em}$\vert$}{0.5em}{}
\titleformat{\subsection}[hang]{\normalsize\bfseries}{\thesubsection}{0.5em}{}
\titleformat{\subsubsection}[hang]{\normalsize\itshape}{\thesubsubsection}{0.5em}{}

\graphicspath{ {./Figures/}}

\begin{document}

\maketitle

\thispagestyle{fancy}
\begingroup
\renewcommand\thefootnote{}\footnotetext{$^{\dagger}$ These authors contributed equally.}
\endgroup

\begin{abstract}
    High-throughput characterization of architected materials across a wide range of length scales enables rapid screening of topologies for engineering applications. Scaled-down specimens manufactured and evaluated in laboratory environments enable this iteration, but application scenarios may involve differing length scales and loading conditions that complicate direct comparisons. Here, we use a spinodal architected morphology to determine the interplay among the constituent material’s strain-rate sensitivity, the topological length scale, and the imposed deformation rates. We report characterization spanning strain rates from $10^{-3}$ s$^{-1}$ to $10^{4}$ s$^{-1}$ on spinodal architected specimens with length scales of 100~\textmu{}m (microscale) and 30~mm (macroscale). The experiments show that while microscale specimens exhibit moderate increase in strength at high strain rates, macroscale specimens exhibit a nearly tenfold increase in strength at equivalent strain rates. Finite element calculations reveal that this increase is linked to a transition from a response governed by constituent material strain-rate sensitivity to inertia-dominated behavior in macroscale specimens---a transition not observed in microscale specimens at the strain rates investigated here. Using extensive finite element calculations, we develop maps to establish the parameters governing the regimes of behavior, illustrating that the transition from behavior governed by constituent material rate sensitivity to inertia-dominated behavior has analogies to fluids in that it depends on a structural length scale. Our findings provide insights into the physical parameters that govern responses across length and time scales, towards the development and design of new laboratory experiments that extract relevant dynamic properties for structural applications.
\end{abstract}

\section{Introduction}

Advancements in additive manufacturing have enabled the fabrication of multifunctional architected materials with tunable properties \cite{Xia_2022,Portela_2020,Jiao_2023}, high strength-to-weight ratios \cite{Schaedler_2011}, programmability \cite{Xia_2022}, and exceptional mechanical behavior \cite{Meza_2015,Jiao_2023,Evans_2010}. Due to their ability to undergo large deformations at nearly constant, tunable stress levels, architected materials are finding applications in energy absorption. For example, they are used in helmets \cite{Murr_2010,Najmon_2018}, to enhance the crashworthiness of automobiles \cite{Fuganti_2000}, in landing gears for space vehicles \cite{Selinanov_2021}, and extensively in blast-resistant structures \cite{Fleck_2004,Liang_2005,McShane_2006,Wei_2008,Rathbun_2006}. In all these applications, the architected materials undergo high-strain-rate and high-amplitude loading, underscoring the importance of understanding their mechanical responses across length and time scales \cite{surjadiEnabling2025}.

Sandwich structureswith a low-density core have a long history \cite{Zenkert_1995} of leveraging various forms of architecture to improve their performance in energy absorption applications. Early sandwich structures had foam cores \cite{Radford_2006}, which then transitioned to more efficient architected topologies. Examples include honeycombs \cite{Rathbun_2006}, Y-shaped structures \cite{Tilbrook_2007} as well as multi-layer pyramidal topologies \cite{Dharmasena_2010} that today are referred to as architected materials. The high-strain-rate or dynamic response of these architected topologies has been extensively investigated \cite{Tilbrook_2007,Dharmasena_2010,Holloman_2014,Radford_HC_2006}. These studies were largely restricted to structures made of metallic materials with low strain-rate sensitivity, and thus the measured rate sensitivities were primarily due to inertia. Two key behavioral regimes were uncovered. In architected topologies where the quasi-static response was stretching-dominated \cite{Deshpande_2001}, micro-inertial effects stabilize local buckling of the unit cells, and this results in a strong strain-rate sensitivity even at low applied velocities/strain rates. This behavior is reminiscent of the Type II structures identified by Calladine and English \cite{Calladine_1984}. Examples of such sandwich cores include the corrugated core \cite{Tilbrook_2007} and honeycomb structures \cite{Fleck_2004}. On the other hand, bending-dominated topologies, such as the Y-core \cite{Tilbrook_2007} and metallic foams \cite{Deshpande_2000}, do not exhibit strong strain-rate sensitivity at low strain rates. This is because micro-inertia has a negligible effect in these cases. In the notation of Calladine and English \cite{Calladine_1984}, these are Type I structures. While differing micro-inertia effects result in a rich range of behaviors at relatively low strain rates/velocities, shock effects dominate in all cases at high velocities, with the topology of the architected material playing a smaller role.

The studies on metallic architected sandwich cores were typically conducted on both centimeter-scale laboratory specimens \cite{Tilbrook_2007,Radford_HC_2006,McShane_Plates_2006,Ferri_2010} and larger-scale structures that underwent field testing \cite{Wei_2008,Dharmasena_2010,Holloman_2014,Dharmsena_2013}. This two-scale testing was motivated by the recognition of structural length-scale effects that arise under dynamic loading conditions. In metallic sandwich structure experiments, these length scale effects were primarily associated with the ratio of the loading time to the structural response time (which scales with the size of the structure) \cite{Fleck_2004}. Length-scale effects associated with the unit-cell size of the architected material topology were largely neglected. Moreover, the studies focused on rate sensitivity resulting from geometry-driven inertial effects. The low rate sensitivity of the constituent metallic materials implied that interactions of material rate sensitivity and inertia were not investigated in detail.

Enabled by advances in additive manufacturing and characterization, recent studies on architected materials have probed their high-rate and impact responses across length scales. At the macroscale, efforts have ranged from drop-weight \cite{Shan2015, Mines2013} to Kolsky-bar experiments \cite{Weeks2022, TancogneDejean2016}, which have identified architecture-dependent deformation mechanisms such as compaction fronts and dynamic responses primarily dominated by material rate-dependence. Leveraging higher-resolution manufacturing techniques, micro- to millimeter-scale samples have enabled the study of shock propagation in these materials, visualized by ultrafast X-ray radiography \cite{hawreliak2016dynamic, dattelbaumShockwave2020,lindDynamicCompression2019}. Motivated by a push for enabling higher-throughput characterization in addition to leveraging size effects of the constituent material, some recent studies have performed dynamic characterization of nano- to microscale architected materials to strain rates of up to 10\textsuperscript{7} s\textsuperscript{-1}. These works have demonstrated high energy absorption efficiency as well as architecture-dependent mechanisms such as dynamic compaction and fracture in the case of microparticle impact testing \cite{Butruille2024, Serles2026}. Due to limitations in the information that can be captured at these small length and time scales, dimensional analysis has been the preferred tool to place these microscale results within an application-relevant context---but several questions remain on the effect of length scale on these dynamic responses.  

Length-scale effects arising from the interaction of viscous forces and inertia dominate fluid mechanics, are captured by non-dimensional groups such as the Reynolds number \cite{Reynolds_1883}. These groups provide the key scaling laws that enable the design of laboratory-scale experiments to appropriately mimic structural applications. With the imminent possibility of fabricating multiscale architected materials with features that range from nanometers to centimeters \cite{Portela_2020, Zheng2016}, questions about length-scale effects in the context of these microstructures have resurfaced. Thus, a search for equivalent non-dimensional groups in architected materials to allow for comparison and extrapolation across a variety of loading conditions and constituent materials has been a topic of recent scientific interest. Notably, across the field, fabrication techniques that allow complex topologies and combine high production volume and high resolution, such as vat photopolymerization \cite{Shaikeea2022} and two-photon lithography \cite{gu2PhotonMetalens2025}, are widely used methods that make extensive use of polymeric constituent materials. These materials are known to exhibit pronounced strain-rate sensitivity; however, the interaction between constituent material strain-rate sensitivity and inertia in architected solids remains poorly understood. Consequently, this implies that an understanding of the non-dimensional groups (akin to Reynolds number) that may be used to appropriately scale laboratory specimens has also not yet been established, but is necessary for their applicability. 

Concurrently, while the properties, deformation mechanisms, and fundamental physics of a wide variety of truss- and plate-based architected designs have been extensively studied, these structures are highly susceptible to manufacturing-induced defects such as strut waviness, wall-thickness variations, and node offsets \cite{Liu_2017,Meza_2017,glaesenerDefects2023}. Such imperfections disrupt the architecture's periodicity and internal symmetry, creating additional stress concentrations and leading to localized damage. As a result, there has been a paradigm shift toward disordered and aperiodic shell-based architectures, particularly spinodal architectures \cite{Portela_2020,Dhulipala_2025,Valdevit_2019}, which eliminate stress concentrations caused by sharp joints and periodicity-breaking defects \cite{Hsieh2019} while also enabling the manufacture of graded structures \cite{Senhora_2022,Kumar_2020, Zheng_2021}. The spinodal decomposition phase-separation process leads to the formation of bicontinuous domains with smooth, doubly curved interfaces with negative Gaussian curvature and near-zero mean curvature \cite{Hsieh2019,Lazarus2012}.  When the interface is thickened to form a continuous shell structure, their inherent Gaussian curvature has been shown to convey numerous mechanical advantages \cite{Dhulipala_2025}, exhibiting stretching-dominated behavior, exceptional stiffness, energy absorption, and recoverability \cite{Hsieh2019,Portela_2020}. These spinodal architectures can, in principle, be realized through self-assembly, suggesting broad applicability comparable to that of beam- or plate-based morphologies with notable mechanical benefits. Owing to these advantages, we selected the spinodal architecture as a representative topology for this study. 

Here we provide an experimental and computational study that attempts to fill the gap in our understanding of the role of length scale and constituent-material rate dependence on the dynamic response of architected materials. We first present an experimental investigation of the compressive response of microscale (micron-scale) and macroscale (millimeter-scale) spinodal architected specimens over a wide range of loading rates, evidencing rather stark differences in the responses across scales. To determine the source of these differences, we employ finite element calculations which highlight an interplay between constituent material strain-rate sensitivity and inertia as the driver behind these differences. Finally, finite element calculations are used to develop maps that identify the regimes of behavior and the associated non-dimensional groups that may be used to design scaled laboratory specimens, and discuss analogies and differences with equivalent scaling in fluid mechanics. Altogether, these efforts combine evidence across several orders of magnitude in length- and time-scales towards establishing an understanding that can guide the design of downscaled experiments as indicators of structural performance at larger length scales. This framework paves the way towards enabling high-throughput characterization and discovery of architected materials via compact and efficient bench-top characterization methods, producing large experimental data sets that, for example, could be used by leading-edge data-driven mechanics frameworks. 
\begin{figure}[htpb]
    \centering
    \includegraphics[width=1\textwidth]{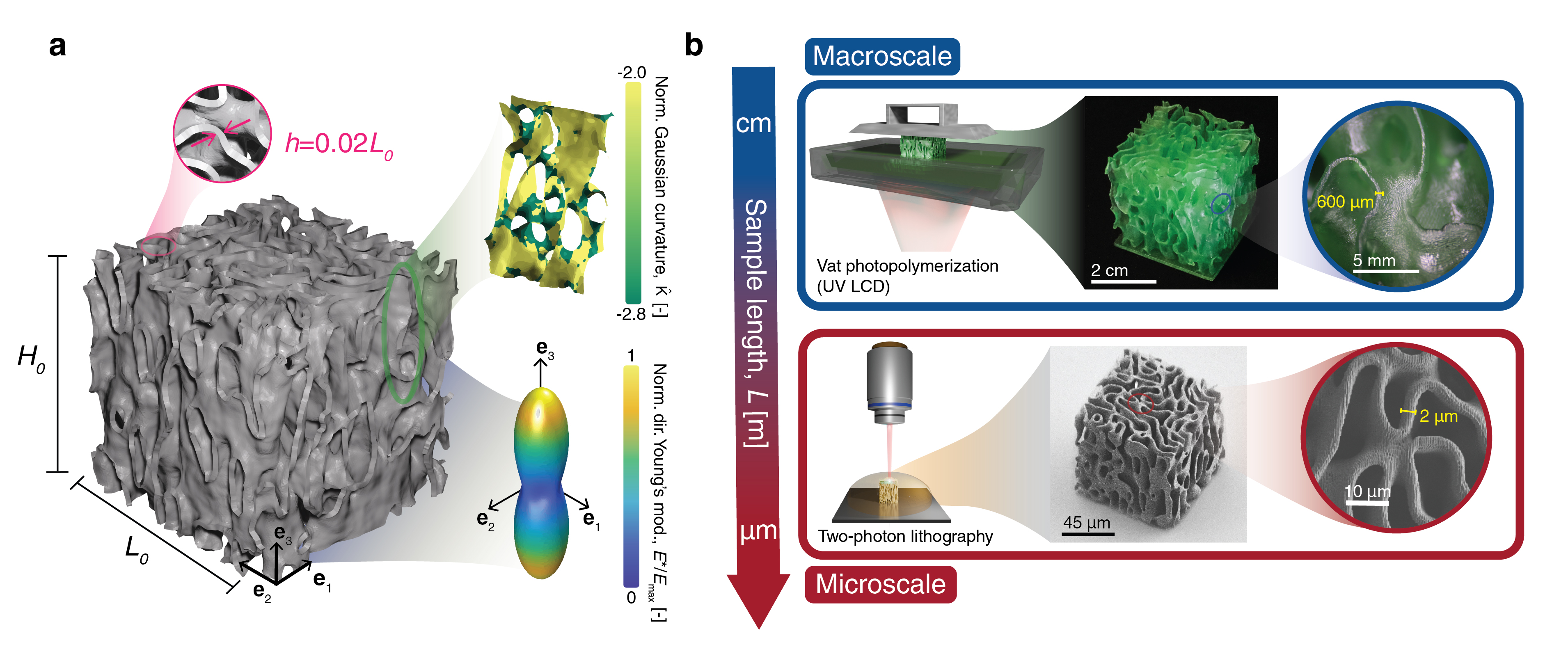}
    \caption{(a) Rendered image of the cubic specimen of side length $L_0$, with insets indicating the relative shell thickness ($h$), normalized Gaussian-curvature distributions ($\hat{K}$), and the homogenized elastic surface of this representative volume element, which evidences highly anisotropic stiffness. (b) Depiction of the macro- (top) and micro-scale (bottom) fabrication routes for the spinodal architected specimens, with insets depicting photographs or micrographs of the manufactured specimens.}
    \label{fig:overview}
\end{figure}

\section{Experimental Protocol}\label{sec:Methods}

We fabricated cuboidal samples of width $L_0$, height $H_0$ and a columnar spinodal architecture (Fig.~\ref{fig:overview}\hyperref[fig:overview]{a}). The specimen topology was generated by modeling a spinodal decomposition process using the Cahn-Hilliard framework \cite{Vidyasagar_2018,Dhulipala_2025}. Anisotropy was introduced via the inclusion of a surface energy term: specifically, the formulation incorporated two preferential directions, $\vectorbold*{e_1}$  and $\vectorbold*{e_2}$, that promote the generation of surfaces with local surface normals along these directions \cite{Vidyasagar_2018}. Since the cube edges were aligned with  the $\vectorbold*{e_i}$ directions, this resulted in a preferential stiffness along the $\vectorbold*{e_3}$ direction. The interfacial surface between the two phases of the spinodal decomposition was extracted to obtain a 3D representative volume element. Geometric analysis of this architecture confirmed that the spinodal morphology possesses primarily low local normalized principal curvatures ($|\hat{\kappa}_1|,|\hat{\kappa}_2|<20$, $\hat{\kappa}_i\equiv\kappa_iL_0$), with, on average, negative normalized Gaussian curvature ($\hat{K}\equiv\hat{\kappa}_1\hat{\kappa}_2$) and near-zero normalized mean curvature ($\hat{M}\equiv0.5(\hat{\kappa}_1+\hat{\kappa}_2)$) in its shells. These surfaces were then thickened until the ratio $h/L_0=0.02$ where $h$ is the nominal shell thickness. With this choice in $h/L_0$, the specimen relative density is $\bar{\rho} \equiv \rho/\rho_s \approx 0.3$, where $\rho$ is the effective density of the specimen and $\rho_s$ is the density of the constituent material. Fabricating these specimens at two length scales, we performed uniaxial compression experiments across $\sim\!{7}$ decades in strain rate (Fig.~\ref{fig:methods}\hyperref[fig:methods]{a}), as outlined below. 

\subsection{Microscale experiments}
\subsubsection{Specimen manufacturing}
The microscale samples were fabricated via two-photon polymerization of IP-Dip2 photoresist (Nanoscribe PPGT2) (Fig.~\ref{fig:overview}\hyperref[fig:overview]{b}). 
The geometry was scaled to have a minimum feature size of approximately 2 \textmu{}m (shell thickness) to ensure sufficiently accurate printing. The specimens were printed with a hatching distance (lateral voxel overlap) of 0.2 \textmu{}m and a slicing distance of 0.3 \textmu{}m. After printing, the samples were developed in propylene glycol methyl ether acetate (PGMEA) for 8 hours, followed by isopropyl alcohol for 8 hours before undergoing critical point drying to prevent damage due to capillary effects (monolithic samples were dried in air). The resulting columnar samples, verified via scanning electron microscopy (SEM), were confirmed to have a relative density of $\bar{\rho}\approx 0.30$ with a height $H_0=83.8\pm1.4$ \textmu{}m, width $L_0=89.8\pm0.6$ \textmu{}m, and a shell thickness $h=2.1\pm0.3$ \textmu{}m (Fig.~\ref{fig:overview}\hyperref[fig:overview]{b}). To determine the properties of the constituent material of the spinodal architected specimens, monolithic pillars of height $H_0=84.5\pm1.8$ \textmu{}m and diameter $d=36.9\pm1.5$ \textmu{}m were also fabricated using the same printing parameters.
\begin{figure}[htpb]
    \centering
    \includegraphics[width=1\textwidth]{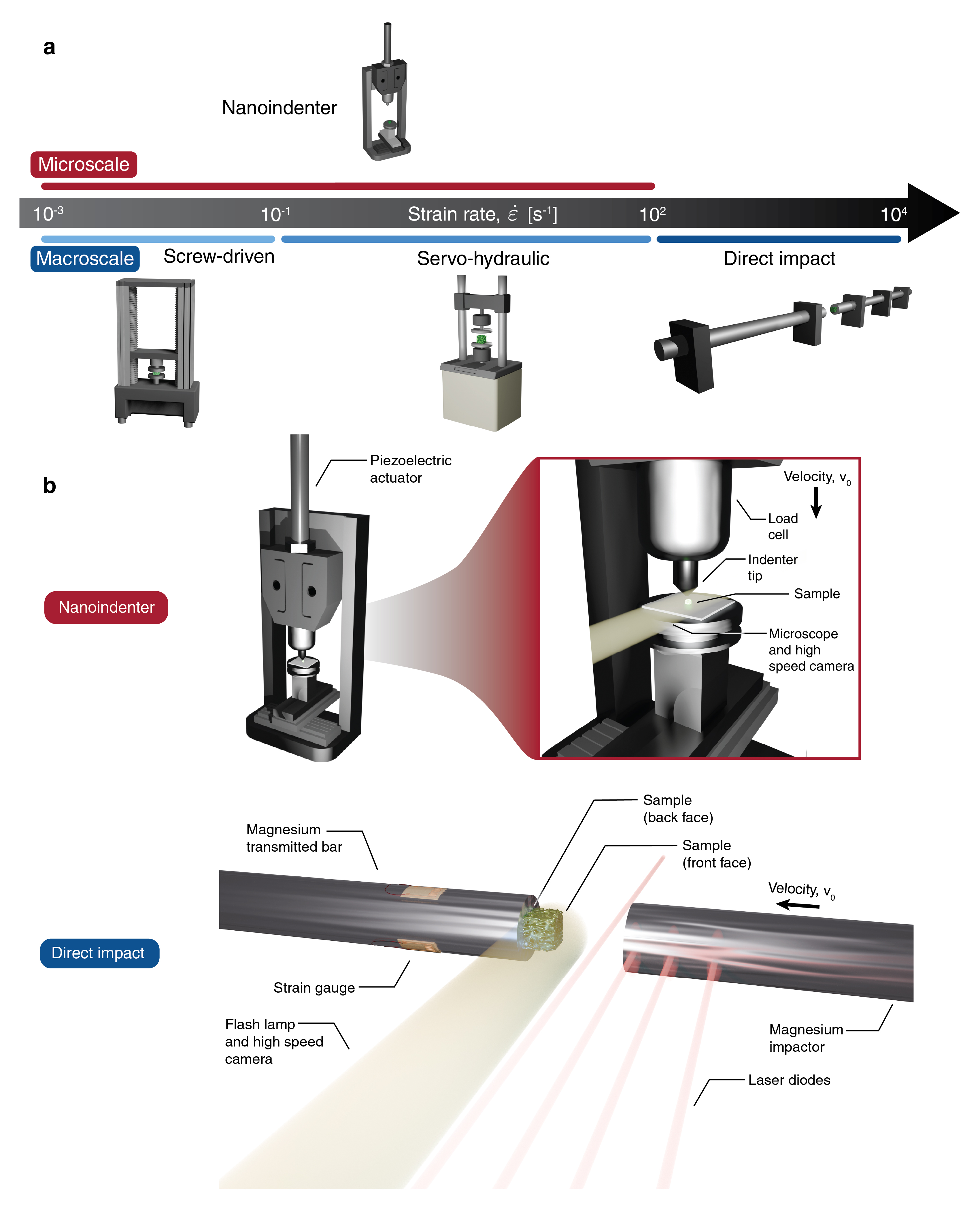}
    \caption{(a) Characterization tools used for the microscale and macroscale compression measurements, together spanning a range of strain rates from 10\textsuperscript{-3} s\textsuperscript{-1} to 10\textsuperscript{4} s\textsuperscript{-1}. (b) Specifics of the nanomechanical setup for microscale measurements (top) and the direct impact setup for macroscale measurements(bottom).}
    \label{fig:methods}
\end{figure}
\subsubsection{Measurement Protocol}

Compression experiments on microscale samples were performed using a displacement-controlled nanoindenter (Alemnis ASA) (Fig.~\ref{fig:methods}\hyperref[fig:methods]{b}), achieving a maximum displacement rate of $v_0=10 \textnormal{ mm} \cdot \textnormal{s}^{-1}$. We conducted experiments over the range $80 \textnormal{ nm} \cdot\textnormal{s}^{-1}\leq v_0\leq 10 \textnormal{ mm}\cdot \textnormal{s}^{-1}$, employing a 400 \textmu{}m diameter diamond flat punch tip. The tip was initially positioned 20 \textmu{}m above the sample’s top surface to allow it to accelerate to the required velocity. A 100 \textmu{}m piezoelectric stack served as the actuator for the tip, while the load was measured using a 2.5 N load cell attached to the displaced tip. Sampling rates ranging from 40 Hz to 100 kHz, depending on the applied strain rate, were used to record measurements from the load cell. These experiments were imaged \textit{in situ} via transmission optical microscopy using incoherent illumination, a $10\times$ microscope objective, and a high-speed camera (Photron FASTCAM Mini AX2000). This setup provided a minimum spatial resolution of 0.35 \textmu{}m/px,  with a maximum frame rate of 10,000 fps, enabling acquisition of a minimum of 100 frames across a loading-unloading cycle.

\subsection{Macroscale experiments}
\subsubsection{Specimen manufacturing}

Cubic macroscale specimens of size $L_0=H_0=30.00 \pm 0.05$ mm were 3D-printed using a
UV LCD printer (Phrozen Sonic Mini 8k) and Anycubic UV Tough resin, allowing a nominal printing resolution of $\sim\!22$ \textmu{}m. This resin was chosen for its high toughness that allows for large deformation of the specimens prior to fracture under high-strain rate conditions. 
To minimize defects, the shell thickness (the minimum feature size) in the printed spinodal architected specimens was $h=620\pm 30$ \textmu{}m which resulted in specimens with a relative density of $\bar{\rho}\approx 0.30$. A layer height of 50 \textmu{}m in the $\vectorbold*{e_3}$-direction and an exposure time of 3 seconds were used during the printing process. Following printing, the samples were initially washed in isopropyl-alcohol (IPA) for 5 minutes and then placed into a clean IPA bath for an additional 5 minutes. The samples were then air-dried for 30 minutes, without a post-curing step, and tested within an hour. Specifically, post-curing was avoided to delay brittle responses of the polymer at high strain rates. The same printing protocol was followed to fabricate 1 cm monolithic cubes of the resin, which were used to determine the properties of the constituent material.

\subsubsection{Measurement protocol}
The manufactured spinodal and monolithic specimens were characterized using three different experimental setups to cover the complete quasi-static to dynamic compression regime (Fig.~\ref{fig:methods}\hyperref[fig:methods]{a}). A screw-driven test machine with a maximum displacement rate of $5 \textnormal{ mm}\cdot \textnormal{s}^{-1}$ was used for the lowest compression-rate experiments. A laser extensometer (recording at 128 Hz) measured the platen displacement while the load cell of the machine measured the applied force. Intermediate strain rates were investigated via a servo-hydraulic test machine capable of imposing displacement rates in the range $3 \textnormal{ mm}\cdot \textnormal{s}^{-1}\leq v_0\leq300 \textnormal{ mm}\cdot \textnormal{s}^{-1}$ on these specimens. The deformation for these experiments was imaged using a high-speed camera (Phantom V12) with a spatial resolution of  63 \textmu{}m/px at frame rates ranging from 400 fps to 3000 fps.

The highest strain rates were probed using a direct impact Kolsky bar setup with a gas gun used to fire the striker rod (Fig.~\ref{fig:methods}\hyperref[fig:methods]{b}). The specimens were attached to a 2 m long AZ31B magnesium bar. Axial strain gauges placed diametrically opposite and 0.5 m from the specimen were wired in a half-bridge arrangement to negate bending of the transmitted bar. The striker rod was fired at the specimen via the gas gun and the force transmitted through the specimen then deduced from the strain gauge readings on the transmitted bar. A series of four laser diodes spaced 20 mm apart provided the impact velocity of the striker rod and a high-speed camera (Phantom V1610) with a spatial resolution of  150 \textmu{}m/px operating at 128,000 fps was used to record the high-rate deformation in the sample. The specimens were speckled using spray paint to provide sufficient contrast during the deformation and illuminated using a flash lamp (Bowens Gemini GM500Pro).

\section{Rate dependence under uniaxial compression} \label{sec:Results}

The compressive responses of the constituent material and architected specimens were measured over a wide range of strain rates for both the microscale and macroscale specimens. We proceed to discuss these measurements while focusing the qualitative differences in the responses of the microscale and macroscale architectures. 

\subsection{Parametrization and interpretation of the measurements}
The compressive response was measured over a wide range of imposed displacement rates: $80 \textnormal{ nm} \cdot\textnormal{s}^{-1}\leq v_0\leq 10 \textnormal{ mm}\cdot \textnormal{s}^{-1}$ for the microscale specimens and $3 \;\textnormal{mm} \cdot\textnormal{s}^{-1}\leq v_0\leq 40 \textnormal{ m}\cdot \textnormal{s}^{-1}$ for the macroscale specimens. Over this wide velocity range there is a possibility that the specimens are not in axial equilibrium and thus the measured response cannot be simply interpreted in terms of a nominal compressive stress versus strain response. We first clarify this interpretation before proceeding to present the measurements.

The inertial stresses scale as $\rho v_0^2$ where $\rho$ is the effective density of the specimen. Under low-rate compression, these inertial stresses $\rho v_0^2$ are significantly smaller than the specimen strength $\sigma_n^Y$. In this case, the specimens are in equilibrium such that the force $F_F$ on the displaced front face of the specimen is equal to the force $F_B$ on the stationary back face of the specimen (Fig.~\ref{fig:methods}\hyperref[fig:methods]{b}). At high compression rates $v_0$, inertial stresses become significant such that the force on the moving front face can exceed the force on the stationary rear face. The force $F_B$ on the stationary rear face is more directly related to the material properties and is of primary interest here. Over the range of $v_0$ investigated here, the constituent material strengths always significantly exceed the inertial stresses, so the monolithic constituent material specimens always remained in axial equilibrium. Thus, in this case the measured forces $F_F$ or $F_B$ are used to define a nominal stress as $\sigma_n\equiv F_F/A_0=F_B/A_0$ where $A_0$ is the cross-sectional area of the undeformed specimen. Further, we can then define a nominal strain rate $\dot{\varepsilon}_n \equiv v_0/H_0$, where $H_0$ is the height of the undeformed specimen and the nominal strain follows as $\varepsilon_n = \int_0^t \dot{\varepsilon}_n dt = v_0 t/H_0$ where time $t=0$ is the instant the specimen deformation commenced.

The behavior of the architected specimens is more complex. Our numerical calculations (Section \ref{sec:Simulation}) show that $\sigma_n^Y \gg \rho v_0^2$ with $F_F=F_B$ for $v_0<10 \textnormal{ m}\cdot \textnormal{s}^{-1}$. Thus, the microscale specimens remain in axial equilibrium over the whole range of responses we report here. On the other hand, in the direct impact experiments, the macroscale specimens may no longer remain in axial equilibrium and a nominal stress within the specimen cannot be defined. Rather, we use the measured force $F_B$ to define a back face stress $\sigma_B \equiv F_B/A_0$ as the stress transmitted through the specimen. Given that $F_F=F_B$ in all other cases, we report all measured stresses for the architected specimens in terms of $\sigma_B$ versus $\varepsilon_n$ which can be interpreted as the force transmitted through the specimen as a function of the imposed displacement.

\begin{figure}[htpb]
    \centering
    \includegraphics[width=1\textwidth]{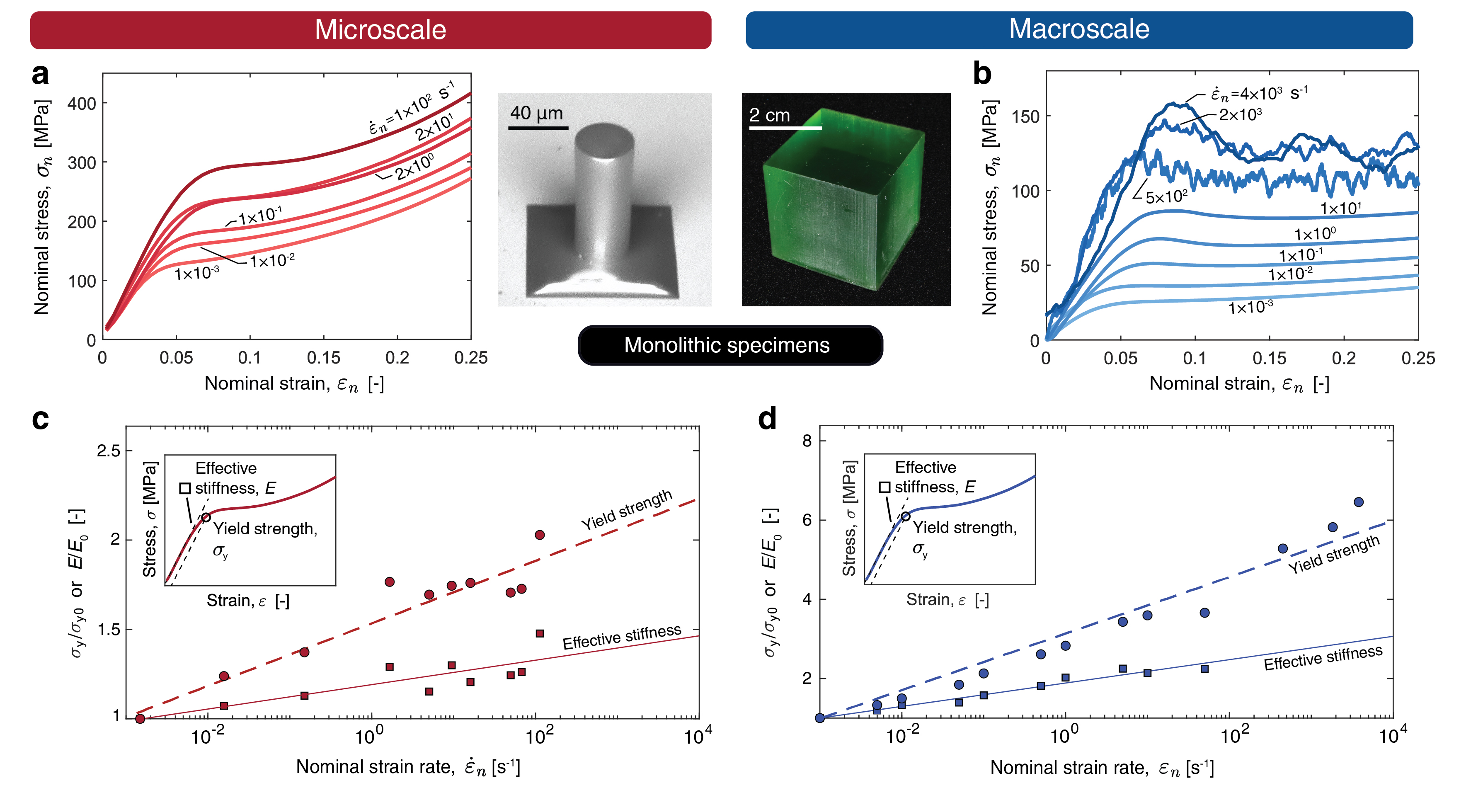}
    \caption{The compressive nominal stress $\sigma_n$ versus strain $\varepsilon_n$ responses of the constituent polymers used to manufacture the (a) micro and (b) macroscale spinodal architected specimens. Summary of the rate dependence of the effective stiffness $E$ and effective yield strength $\sigma_Y$ (defined in the insets) as a function of strain rate $\dot{\varepsilon}_n$ for the constituent materials used to manufacture the (c) micro and (d) macroscale spinodal architected specimens. The effective stiffness and strength are normalized by their quasi-static values at $\dot{\varepsilon}_n=10^{-3}$~s$^{-1}$  and denoted by $E_0$ and $\sigma_{Y0}$, respectively.}
    \label{fig:basematerial}
\end{figure}

\subsection{Constituent Material Characterization}
The constituent materials used to manufacture the microscale and macroscale spinodal architected specimens are rate-sensitive amorphous polymers, exhibiting both viscoelastic and viscoplastic properties. While it would have been ideal to utilize the same polymer at both length scales, this was not possible as the micro- (micron scale) and macro- (millimeter scale) fabrication processes require different crosslinking wavelengths in the polymers. To establish the response of the constituent materials over the range of strain rates of interest, uniaxial compression experiments were carried out on the monolithic specimens of these polymers manufactured using the same route as that used for the spinodal specimens. The measured $\sigma_n$ versus $\varepsilon_n$ curves, included in Figs.~\ref{fig:basematerial}\hyperref[fig:basematerial]{a} and ~\ref{fig:basematerial}\hyperref[fig:basematerial]{b}, suggest that the behavior of the polymers used for the microscale and macroscale specimens is similar, with both polymers displaying a rate-dependent elasto-plastic response. A qualitative difference between the two polymers is the higher strain-hardening of the polymer used in the microscale specimens. To quantify the difference in their mechanical behavior, we define an effective elastic stiffness $E$ as the initial linear slope of the measured $\sigma_n-\varepsilon_n$ curves and the effective yield strength $\sigma_Y$ as the 0.2\% offset stress of the $\sigma_n-\varepsilon_n$ curves (see insets of Figs.~\ref{fig:basematerial}\hyperref[fig:basematerial]{c,d}). With the stiffness and yield strength at the lowest strain rate of $\dot{\varepsilon}_n = 10^{-3}$ s$^{-1}$ denoted as $E_0$ and $\sigma_{Y0}$, respectively, we examine the strain-rate dependency via the dependence of $E/E_0$ and $\sigma_Y/\sigma_{Y0}$ on $\dot{\varepsilon}_n$, as shown in Figs.~\ref{fig:basematerial}\hyperref[fig:basematerial]{c,d} for the polymers used in the microscale and macroscale specimens, respectively. For both resins the effective stiffness and the effective yield strength increase with increasing strain rate. However, the rate at which the yield strength increases with strain rate is higher than that of the modulus, also leading to the yield strain ($\sigma_Y/E$) increasing with strain rate for both polymers.

\begin{figure}[htpb]
    \centering
    \includegraphics[width=1\textwidth]{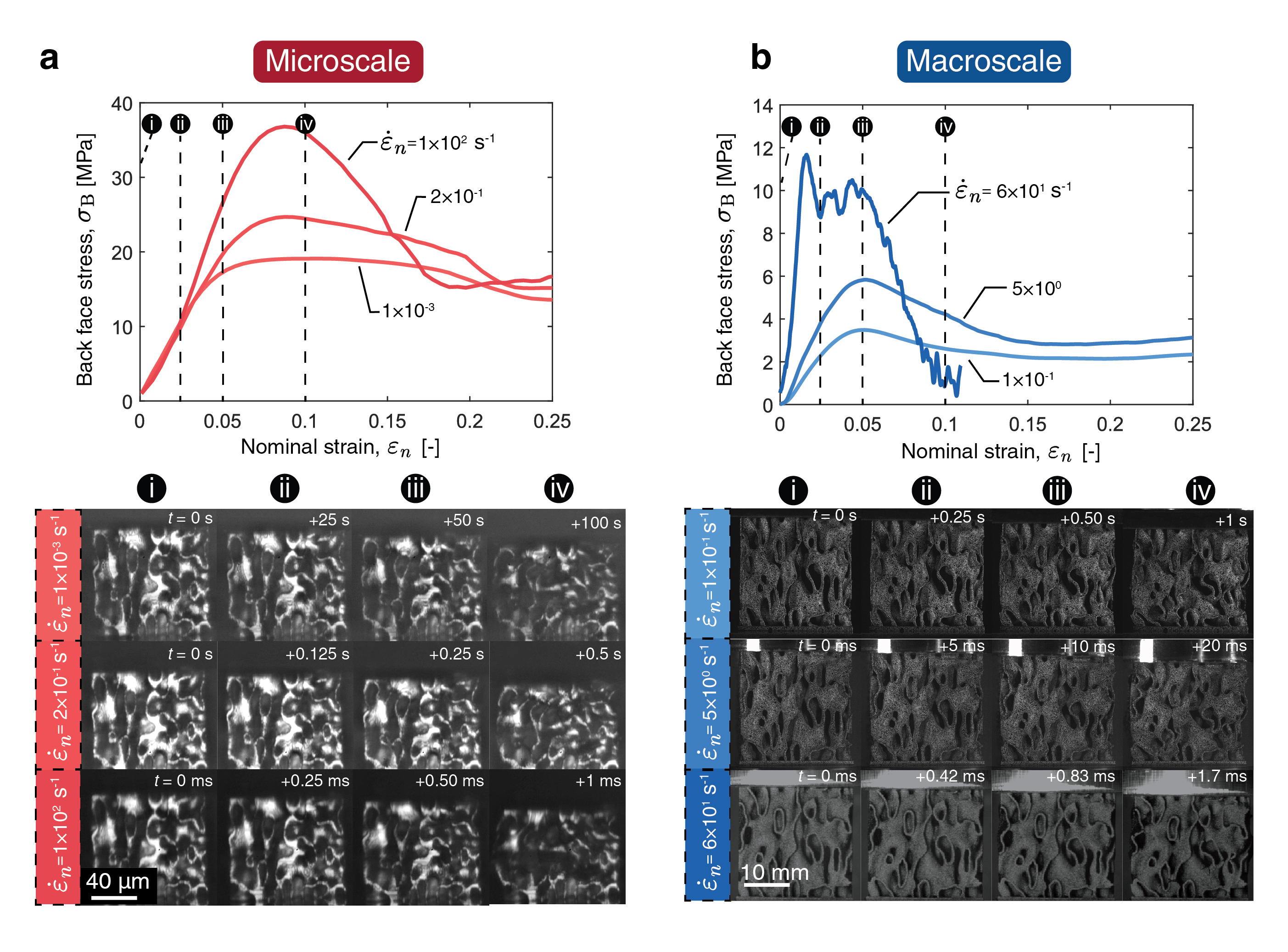}
    \caption{A comparison between the deformation modes of the (a) microscale and (b) macroscale spinodal architected specimens for three imposed strain rates spanning five orders of magnitude. The corresponding measured stress versus strain responses are included and key stages of deformation are marked corresponding to the images.}
    \label{fig:insitu}
\end{figure}

\subsection{Compressive response of the spinodal specimens}
\begin{figure}[htpb]
    \centering
    \includegraphics[width=1\textwidth]{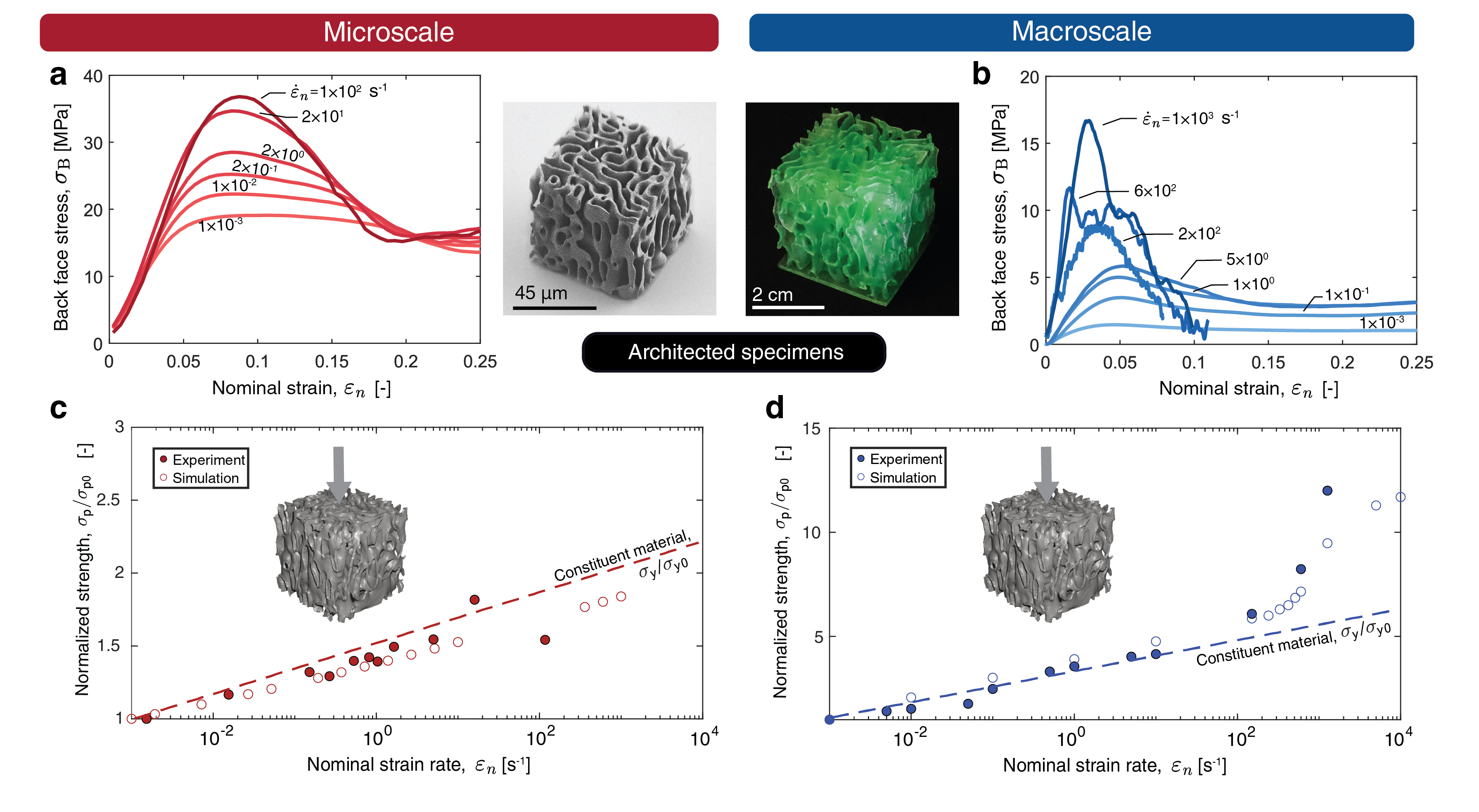}
    \caption{The back face stress $\sigma_B$ versus strain $\varepsilon_n$ responses of the (a) micro and (b) macroscale spinodal architected specimens. Summary of the rate dependence peak strength $\sigma_p$ as a function of strain rate $\dot{\varepsilon}_n$ for the (c) micro and (d) macroscale spinodal architected specimens. The peak strength is normalized by its corresponding quasi-static value $\sigma_{p0}$  at $\dot{\varepsilon}_n=10^{-3}$  s$^{-1}$. Corresponding FE predictions are also included in (c) and (d). The dashed lines show the evolution of strain-rate dependency of the constituent materials, expressed as $\sigma_Y/\sigma_{Y0}$ versus $\dot{\varepsilon}_n$, reproduced from Figs.~\ref{fig:basematerial}c,d for the micro and macroscale specimens, respectively.}
    \label{fig:architecture}
\end{figure}

Compression experiments over the same range of nominal strain rates were performed on the microscale and macroscale spinodal architected specimens. Three representative measurements of the back face stress $\sigma_B$ versus $\varepsilon_n$ responses are included in Figs.~\ref{fig:insitu}\hyperref[fig:insitu]{a,b} for the microscale and macroscale specimens, respectively, for strain rates $\dot{\varepsilon}_n$ ranging from 0.1 s$^{-1}$ to 600 s$^{-1}$. From image frames of the observed deformations (Figs.~\ref{fig:insitu}\hyperref[fig:insitu]{a,b}), with time $t=0$ defined as the instant at which the deformation of the specimen initiated, no clear differences in the deformation modes were observed across length or time scales. Moreover, the architected specimens display rate-dependent behavior qualitatively similar to that of the constituent materials (Figs.~\ref{fig:basematerial}\hyperref[fig:basematerial]{a,b}), viz. both the modulus and strength increase with increasing strain rate. However, unlike the constituent materials, the architected specimens soften after yield at a rate that increases with increasing strain rate, especially for the macroscale specimens. Such softening responses of architected materials have been extensively reported \cite{Calladine_1984,Tilbrook_2007} and usually attributed to the local buckling of architected topologies. This buckling is known to result in rate dependence associated with inertial stabilization against buckling; so-called type II behavior \cite{Calladine_1984}. A more complete set of $\sigma_B-\varepsilon_n$ curves over a wide range of rates $\dot{\varepsilon}_n$ are included in Figs.~\ref{fig:architecture}\hyperref[fig:architecture]{a,b} for the microscale and macroscale specimens, respectively. These measurements and the observations of the deformation modes in Figs.~\ref{fig:insitu}\hyperref[fig:insitu]{a,b} are insufficient to separate strain-rate dependence resulting from inertial effects and the inherent constituent material rate dependence.

To de-convolute the role of constituent material rate dependence from possible inertial effects we include in Figs.~\ref{fig:architecture}\hyperref[fig:architecture]{c,d} the variation of $\sigma_p/\sigma_{p0}$ with $\dot{\varepsilon}_n$ for the microscale and macroscale specimens, where $\sigma_p$ is the peak back face stress of the architected material specimen, and $\sigma_{p0}$ is the value of $\sigma_p$ at $\dot{\varepsilon}_n = 10^{-3}$  s$^{-1}$. Also plotting corresponding variations of $\sigma_Y/\sigma_{Y0}$ versus $\dot{\varepsilon}_n$ for the microscale and macroscale monolithic specimens in Figs.~\ref{fig:architecture}\hyperref[fig:architecture]{c,d} allows a direct comparison between the constituent material and architected specimen rate sensitivities and reveal contrasting behaviors of the microscale and macroscale specimens. Namely, the rate dependence of the microscale specimens is well approximated by the constituent material rate dependence over the whole range of strain rates investigated, viz. $10^{-3}  \textnormal{ s}^{-1}\leq \dot{\varepsilon}_n \leq 10^2  \textnormal{ s}^{-1}$. This suggests that dependence of the strength of the microscale architected specimens is dominated by the constituent material strain-rate dependence. By contrast, the macroscale specimens display a more complex response. For strain rates $\dot{\varepsilon}_n \lesssim 10^2 \textnormal{ s}^{-1}$ the yield strength of the architected specimens again follows the constituent material strain-rate dependence. However, for $\dot{\varepsilon}_n \gtrsim 10^2  \textnormal{ s}^{-1}$ the behavior undergoes a transition with $\sigma_p$ increasing sharply with $\dot{\varepsilon}_n$, i.e., the yield strength of the architected specimens shows a stronger rate dependence compared to the constituent material response.
Due to limitations of the experimental equipment at microscale, which limited experiments to rates of $\dot{\varepsilon}_n \lesssim 10^2  \textnormal{ s}^{-1}$, it is unclear whether a similar transition in behavior will also exist in the microscale specimens at $\dot{\varepsilon}_n > 10^2 \textnormal{ s}^{-1}$ or whether this transition in Fig.~\ref{fig:architecture}\hyperref[fig:architecture]{d} is associated with the specimen length scale. To develop a further understanding of this observed transition, we proceeded to employ simulations as a way to probe the deformation mechanisms independently.

\section{Computational investigation of length-scale dependent responses} \label{sec:Simulation}

To develop an understanding of the differences in the measured responses of the microscale and macroscale architected material specimens, we performed finite element (FE) calculations of the specimens at both length scales. All calculations were performed using ABAQUS/Explicit \cite{Abaqus_manual}, where spinodal morphologies were discretized using 3-node reduced integration shell elements (S3R). The compressive loading was imposed by fixing all degrees of freedom (rotation and translation) on the back surface of the specimen and imposing an axial velocity $v_0$ to the front-face nodes with other degrees of freedom left free. The front-face force $F_F$ and back-face force $F_B$ as a function of time were exported from these calculations.

\subsection{Predictions of the responses of the microscale and macroscale specimens}\label{sec:SimulationOrig}

Prior to investigating the source of any differences in the macroscale and microscale specimen responses, we first establish the fidelity of the FE calculations in capturing the measured behaviors. We model the constituent material as an isotropic elastic perfectly plastic solid with Young’s modulus $E$, Poisson's ratio $\nu_s=0.3$, yield strength $\sigma_Y$ and density $\rho_s=1400$ or $1100 \textnormal{ kg}\cdot \textnormal{m}^{-3}$ for the microscale or macroscale specimens, respectively. Recall that the architected specimen and constituent material strain rate sensitivities follow a very similar trend (Figs.~\ref{fig:architecture}\hyperref[fig:architecture]{c,d}), at least for strain rates $\dot{\varepsilon}_n \lesssim 10^2  \textnormal{ s}^{-1}$, suggesting that the on-average microscopic strain rates for this spinodal architecture approximate the macroscopic applied strain rates. Using this observation, we simplify the calculations by taking the constituent materials to be rate independent for each simulation but include rate dependence in an approximate manner by assuming that the imposed nominal strain rate $\dot{\varepsilon}_n$ on the architected specimen is equal to the local constituent material strain rate $\dot{\varepsilon}$. Thus, for a given imposed velocity, $v_0$, and the equivalent imposed strain rate, $\dot{\varepsilon}_n\equiv v_0/H_0$, we set the constituent material strength, $\sigma_Y$, and modulus, $E$, to the corresponding values given in Figs.~\ref{fig:basematerial}\hyperref[fig:basematerial]{c,d} for the microscale and macroscale specimens, respectively. 

In line with the experiments, the predicted back face force $F_B$ is used to define the back face stress $\sigma_B$ and thereby determine the peak back face strength $\sigma_p$. Comparison of the measured and predicted normalized peak strengths $\sigma_p/\sigma_{p0}$ are included in Figs.~\ref{fig:architecture}\hyperref[fig:architecture]{c,d} for the microscale and macroscale specimens, respectively, over strain rates in the range $10^{-3}  \textnormal{ s}^{-1}\leq \dot{\varepsilon}_n \leq 10^3  \textnormal{ s}^{-1}$. Excellent agreement between the experiments and simulations confirms the fidelity of the FE model. Most interestingly, the FE predictions capture the measured upturn in $\sigma_p$ in the macroscale specimens at $\dot{\varepsilon}_n\sim 10^2$ s$^{-1}$. On the other hand, the FE calculations predict that the peak strength of the microscale specimens follow the rate-dependent yield strength scaling of the constituent material to strain rates of at least $\dot{\varepsilon}_n\sim 10^3$ s$^{-1}$. The calculations thus clearly show that the strain-rate sensitivity of the architected specimen response is length scale dependent.

The source of this difference can be understood by recalling that for $\dot{\varepsilon}_n\sim 10^2$ s$^{-1}$, the microscale and macroscale specimens have imposed velocities $v_0=8 \textnormal{ mm}\cdot \textnormal{s}^{-1}$ and $3 \textnormal{ m}\cdot \textnormal{s}^{-1}$, respectively. Since the inertial stresses scale with $\rho v_0^2$, the upturn in the macroscale behavior is likely to be associated with inertial effects which are absent in the microscale specimens due to the significantly lower deformation velocities. In fact, we note that the FE calculations show that the front and back face forces begin to diverge from each other for $v_0>10 \textnormal{ m}\cdot \textnormal{s}^{-1}$. A velocity  $v_0=10 \textnormal{ m}\cdot \textnormal{s}^{-1}$ corresponds to $\dot{\varepsilon}_n \sim 330  \textnormal{ s}^{-1}$ for the macroscale specimens, i.e., the strain rate at which the transition in the behavior is seen in Fig.~\ref{fig:architecture}\hyperref[fig:architecture]{d}. This strongly suggests that the upturn in the strength of the macroscale specimens is associated with inertial effects.

\subsection{Investigation of the interplay of rate sensitivity, inertia, and length scale}
\begin{figure}[htpb]
    \centering
    \includegraphics[width=1\textwidth]{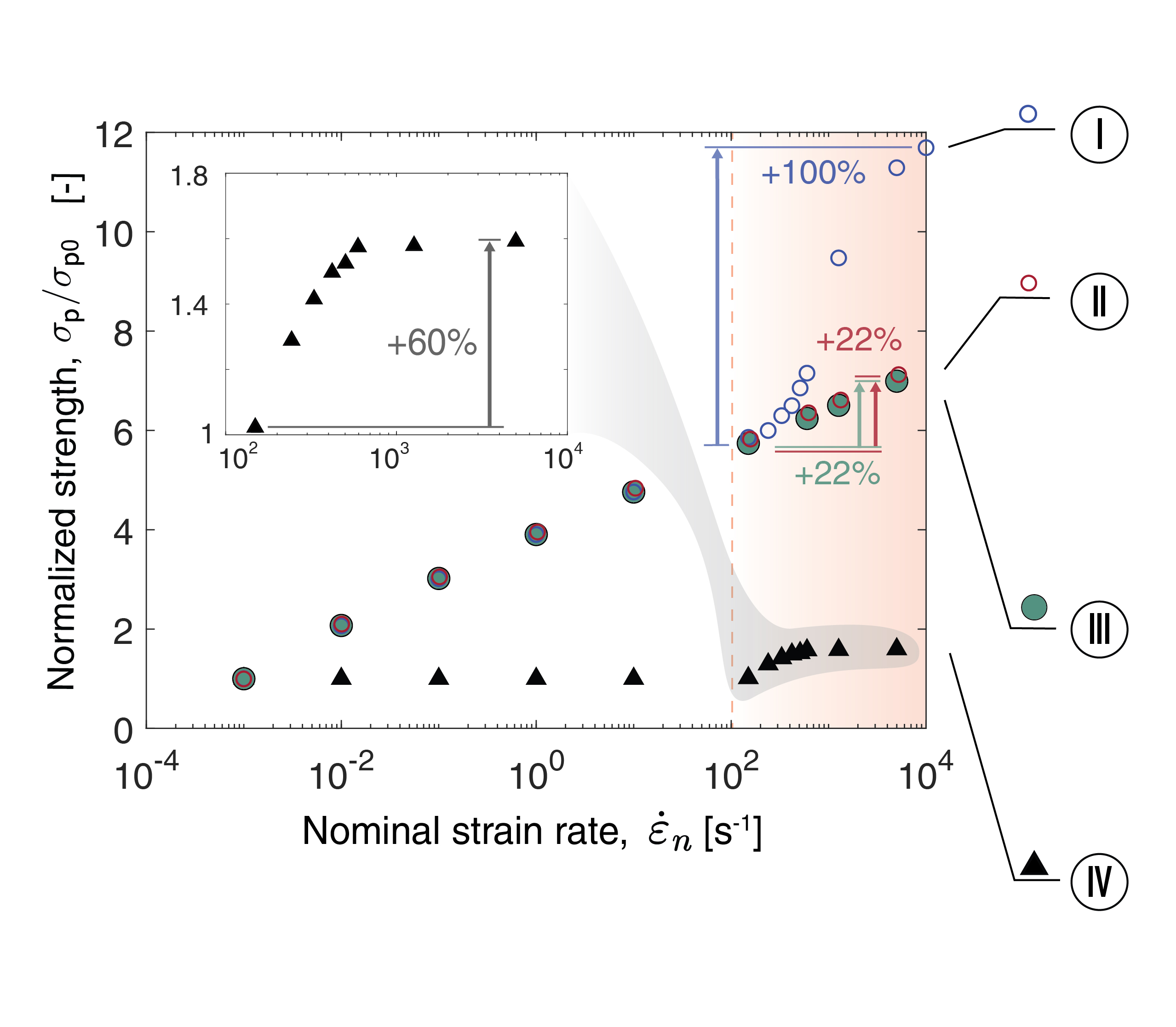}
    \caption{FE calculations to evaluate the interplay of constituent material rate sensitivity, inertia and specimen length scale in setting the response of the architected specimens. Four sets of calculations are reported. Case I: reference case of the macroscale specimen, Case II: Effect of length scale, Case III: Effect of isolating constituent material rate sensitivity, and Case IV: Effect of isolating inertia.}
    \label{fig:FEStudy}
\end{figure}
The above FE calculations pointed to the upturn in the strength of the macroscale specimens being due to inertial effects, while the microscale specimens are dominated by the strain-rate sensitivity of the constituent material. However, the microscale and macroscale specimens used in the experimental study (and therefore also the FE calculations of Section \ref{sec:SimulationOrig}) used different constituent materials. 
To clarify the interplay of constituent material rate sensitivity, inertia, and specimen length scale, we use additional FE calculations where each mechanism was probed independently. 

We consider four sets of calculations to decouple the effects of constituent material rate sensitivity, inertia, and specimen length scale:
\begin{table}[H]
\renewcommand{\arraystretch}{1.25}
\centering
\begin{tabular}{l p{14cm}}

(Case I) & Reference: The macroscale specimen calculations of section \ref{sec:SimulationOrig}, which include the effects of inertia and constituent material rate sensitivity for the macroscale specimens.\\ 

(Case II) & Length scale effect: The calculations as in the reference (Case I) but the specimen self-similarly scaled down to the microscale specimen size. Since these calculations include the effects of inertia and constituent material rate sensitivity of the macroscale constituent material, a comparison with the reference (Case I) identifies the effect of specimen length scale.\\ 

(Case III) & Constituent material rate-sensitivity effect: The macroscale specimen calculations with the rate-dependent constituent material properties of the macroscale specimens as in the reference (Case I). However, all calculations performed at an imposed strain rate of $\dot{\varepsilon}_n = 10^{-3}  \textnormal{ s}^{-1}$---such that inertial effects are negligible in these calculations. However, the material model appropriately accounts for rate dependence.\\ 

(Case IV) & Inertia effect: The macroscale specimen calculations as in the reference (Case I) but the constituent material properties fixed at their values for an imposed strain rate of $\dot{\varepsilon}_n = 10^{-3}  \textnormal{ s}^{-1}$, only triggering inertial effects across strain rates.\\ 

\end{tabular}
\label{tab:FeStudy}
\end{table}
The predicted normalized peak back-face stresses $\sigma_p/\sigma_{p0}$ from these four sets of calculations as a function of $\dot{\varepsilon}_n$ are included in Fig.~\ref{fig:FEStudy}.  In each case $\sigma_{p0}$ is the value of $\sigma_p$ at $\dot{\varepsilon}_n = 10^{-3}  \textnormal{ s}^{-1}$. We now consider each of these calculations in turn. When the specimen is scaled down (Case II), the peak stress follows the reference case for $\dot{\varepsilon}_n \lesssim 10^{2}  \textnormal{ s}^{-1}$. However, while there is an upturn in the reference (Case I) $\sigma_p$ at higher strain rates, no such upturn is observed for the scaled-down specimens. This clearly shows that the upturn seen in the macroscale specimen measurements is associated with the larger specimen sizes. To identify the reason for the upturn in the macroscale specimens, we proceeded to only account for material rate dependence (Case III). The predictions of Cases II and III are nearly indistinguishable, and both show no upturn in $\sigma_p$. Evidence from this case demonstrates that the upturn is caused by inertia and not constituent material rate sensitivity. Therefore, the fact that Cases II and III give nearly identical responses shows that the microscale specimen responses are dominated by constituent material rate sensitivity with little influence of inertia. Finally, eliminating constituent material rate sensitivity and solely focusing on inertia (Case IV) results in a response that is rate-independent for strain rates $\dot{\varepsilon}_n \lesssim 10^{2}  \textnormal{ s}^{-1}$ but evidences a strong relative upturn in the peak strength for $\dot{\varepsilon}_n \gtrsim 10^{2}  \textnormal{ s}^{-1}$. The fractional increase in strength in Case IV for $\dot{\varepsilon}_n \gtrsim 10^{2}  \textnormal{ s}^{-1}$ is similar to the relative increase in the reference case, reinforcing the fact that inertial effects begin to play an important role in the macroscale specimens for $\dot{\varepsilon}_n \gtrsim 10^{2}  \textnormal{ s}^{-1}$. We note that the relative increase at large strain rates for Cases I and IV is not identical. While primarily dominated by inertia, the discrepancy arises from material rate sensitivity and potential coupling with inertial effects.

\section{Deformation mechanism map for the rate dependent response} \label{sec:Inertia}

The calculations in Section \ref{sec:Simulation} have clarified that, while the response on the macroscale architected specimens is constituent-material strain-rate sensitivity dominated at low rates, inertial effects become significant at higher rates. Conversely, over the same range of applied strain rates, inertial effects play little role in the response of microscale specimens with their behavior governed solely by constituent material rate sensitivity. We now proceed to quantify the regimes of dominance of inertia and constituent material rate sensitivity and discuss how these regimes are influenced by specimen size.

The aim here is to develop maps to illustrate the regimes of behavior rather than focus on specific constituent materials. Keeping this in mind, we consider a generic rate-dependent $J_2$-flow-theory material of density $\rho_s$ with a rate independent Young’s modulus $E$ and Poisson ratio $\nu$. The constituent material has no strain hardening and the strain-rate dependence is included only in the yield strength via an overstress model such that the uniaxial tensile yield strength $Y$ of the material is related to the uniaxial tensile strain rate $\dot{\varepsilon}$ via 

\begin{equation}
Y = \begin{cases}
        \sigma_0 & \dot{\varepsilon}\leq \dot{\varepsilon}_0 \\
        \sigma_0 + \alpha \left[\left(\frac{\dot{\varepsilon}}{\dot{\varepsilon}_0}\right)^N - 1\right] & \dot{\varepsilon}> \dot{\varepsilon}_0 
    \end{cases}
    ,
    \label{eq:InertiaMatProp}
\end{equation}
where $\sigma_0$ is the static strength of the solid material, $\alpha$ is a reference strength that scales the rate sensitivity, $N$ is the strain-rate sensitivity exponent, and $\dot{\varepsilon}_0$ is a reference strain rate. In the limit $N\rightarrow0$, the material exhibits no strain-rate sensitivity, and $Y=\sigma_0$. Given this generic and simple constituent material rate sensitivity, dimensional analysis dictates that the back face stress $\sigma_B$ for an architected specimen of density $\rho = \rho_s \bar{\rho}$ is given by
\begin{equation}
    \frac{\sigma_B}{\sigma_0} = \mathcal{G}\left(\frac{\rho v_0^2}{\sigma_0}, \frac{v_0}{H_0\dot{\varepsilon}_0}, \frac{\alpha}{\sigma_0}, N, \bar{\rho}\right),
\end{equation}
where $H_0$ is a representative length scale here taken to be the specimen height. The non-dimensional groups $\alpha/\sigma_0$, $N$, $\bar{\rho}$ are material groups independent of loading while $\rho v_0^2/\sigma_0$, and $v_0/(H_0 \dot{\varepsilon}_0)$ describe the imposed loading in terms of inertia and constituent material rate sensitivity, respectively. Our aim is to quantify the regimes of behavior as a function of the loading and all calculations reported here use the $\bar{\rho} = 0.3$ spinodal architecture considered throughout this study and constituent material properties listed in Table~\ref{tab:MapProps}.

\begin{table}[H]
{\renewcommand{\arraystretch}{1.2}
\begin{tabular}{ccccccc} \hline \hline
$E$ {[}MPa{]} & $\sigma_0$ {[}MPa{]} & $\alpha$ {[}MPa{]} & $N$ {[}-{]} & $\dot{\varepsilon}_0$ {[}s$^{-1}${]} & $\rho_0$ {[}kg/m$^3${]} & $\bar{\rho}$ {[}-{]} \\ \hline
1000          & 10                   & 1                  & 0.3         & $1\times 10^{-3}$                    & 1100                    & 0.3                 \\ \hline\hline
\end{tabular}}
\caption{Material parameters of the constituent solid used to obtain the inertia map.}
\label{tab:MapProps}
\end{table}
\begin{figure}[htpb]
    \centering
    \includegraphics[width=1\textwidth]{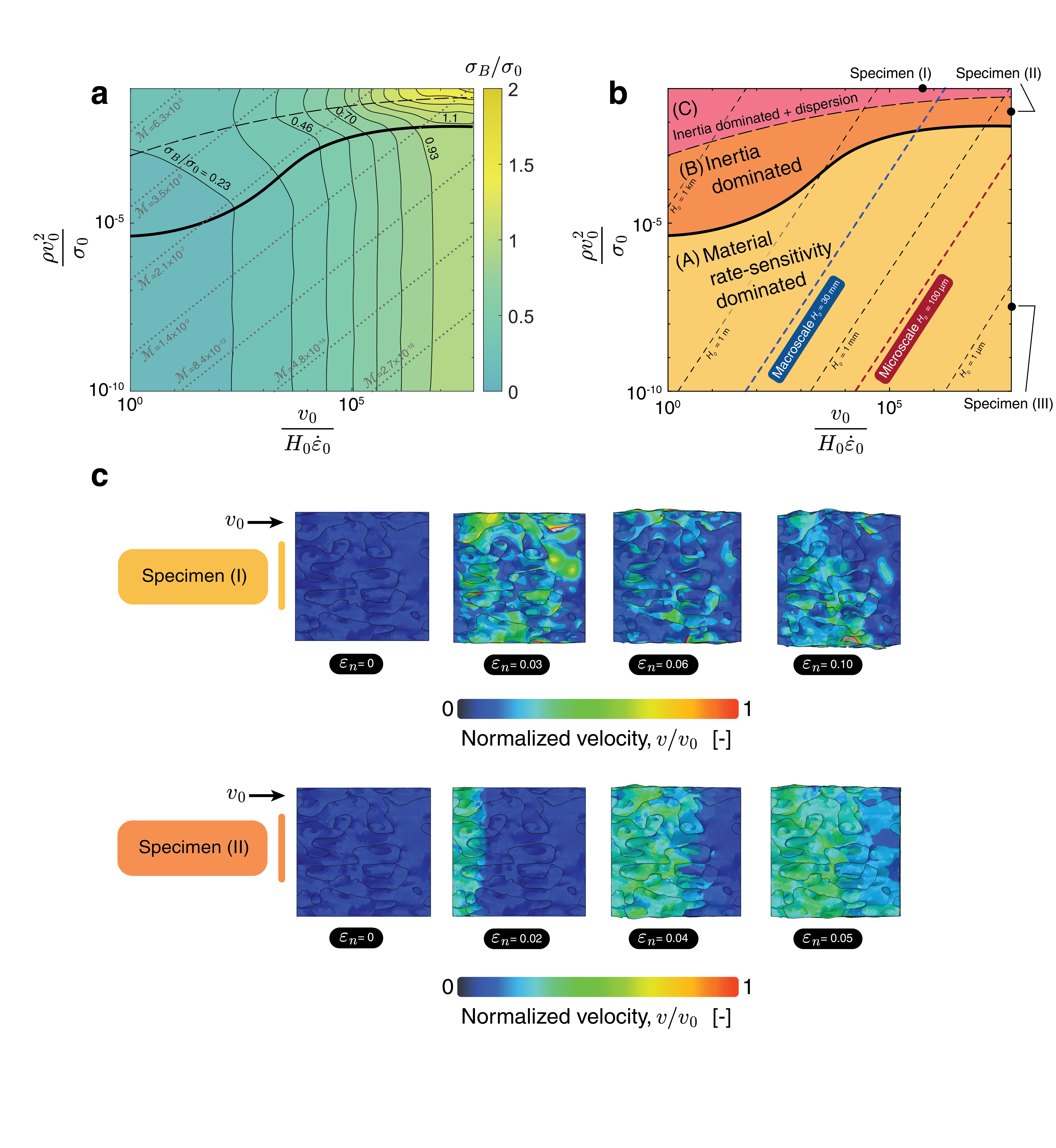}
    \caption{(a) Map with contours of the normalized back face stresses $\sigma_B/\sigma_0$ of the $\bar{\rho}=0.3$ architected specimens. The map has axes $\rho v_0^2/\sigma_0$ and $v_0/(H_0\dot{\varepsilon}_0)$ to parameterize the inertial and viscous forces, respectively. (b) Depiction of the regimes in the map shown in (a), namely: (A) constituent-material strain-rate sensitivity governed response, (B) inertia-governed response and (C) inertia plus dispersion-governed response. In (a) we also include contours of the non-dimensional group $\mathcal{M}$ that is akin to a Reynolds number, while in (b) we include contours of $H_0$ for the material length scale $l=174$ mm based on the material parameters in Table 1. (c) Simulations of the spatial distributions of the normalized material-point velocities $v/v_0$ for specimens (I) and (II), at selected values of nominal strain $\varepsilon_n\equiv v_0 t/H_0$.}
    \label{fig:dimensionlessmap}
\end{figure}
\subsection{Deformation mechanism map}
Our aim is to develop an understanding of the dependence on the normalized back face stress $\sigma_B/\sigma_0$ as a function of the two loading parameters $\rho v_0^2/\sigma_0$, and $v_0/(H_0 \dot{\varepsilon}_0)$. With this in mind, and the specimen fixed to be the $\bar{\rho} = 0.3$ spinodal architecture, we conducted an array of FE calculations employing the protocol outlined in Section~\ref{sec:Simulation}. The two parameters were controlled by self-similarly scaling the specimen by varying its height $H_0$ and/or changing the loading velocity $v_0$. Approximately 200 calculations were conducted to over the ranges $10^{-15}\leq \rho v_0^2/\sigma_0 \leq 1$ and $10^{-7}\leq v_0/(H_0 \dot{\varepsilon}_0) \leq10^{13}$ and the key findings are summarized in Fig.~\ref{fig:dimensionlessmap}\hyperref[fig:dimensionlessmap]{a} in the form of a map with axes $v_0/(H_0 \dot{\varepsilon}_0)$ and $\rho v_0^2/\sigma_0$ on logarithmic scales and contours of $\sigma_B/\sigma_0$.

Before discussing the findings, let us consider the two limits of responses: (i) constituent-material strain-rate sensitivity dominated, and (ii) inertia-dominated. If the response is solely governed by the constituent-material strain-rate sensitivity, $\sigma_B/\sigma_0$ is independent of $\rho v_0^2/\sigma_0$ and thus contours of $\sigma_B/\sigma_0$ will be vertical lines on the map with abscissa of $v_0/(H_0 \dot{\varepsilon}_0)$ and ordinate of $\rho v_0^2/\sigma_0$. Similarly, if the response is only inertia governed, contours of $\sigma_B/\sigma_0$ will be horizontal lines on this map. These two limits are clearly visible on the map (Fig.~\ref{fig:dimensionlessmap}\hyperref[fig:dimensionlessmap]{a}): for low values of $\rho v_0^2/\sigma_0$ we observe that the contours of $\sigma_B/\sigma_0$ are vertical lines, while increasing $\rho v_0^2/\sigma_0$ results in the contours turning and becoming more horizontal, most clearly seen at the top right of the map. 

Using the criterion that the bending of the vertical contours of $\sigma_B/\sigma_0$ represents the transition from constituent-material strain-rate sensitivity governed response to an inertia-governed response, we approximately mark that transition in Fig.~\ref{fig:dimensionlessmap}\hyperref[fig:dimensionlessmap]{a}, and depict these domains in Fig.~\ref{fig:dimensionlessmap}\hyperref[fig:dimensionlessmap]{b}. The qualitative differences in the responses in these two regimes can be visualized by examining the distribution of the magnitude $v$ of the material-point velocity within the specimen as a function of $\varepsilon_n\equiv v_0 t/H_0$. Consider two specimens at locations (I) and (II) in Fig.~\ref{fig:dimensionlessmap}\hyperref[fig:dimensionlessmap]{b}, representing loading in the constituent-material strain-rate sensitivity governed and inertia-governed responses, respectively. Spatial distributions of the magnitude $v$ of the material-point velocity normalized by the imposed velocity $v_0$ are included in Fig.~\ref{fig:dimensionlessmap}\hyperref[fig:dimensionlessmap]{c} for selected values of the nominal strain $\varepsilon_n$. We clearly observe that movement is more localized near the displaced end of the specimen in the inertia-dominated regime (specimen II) and more evenly distributed in the constituent-material strain-rate sensitivity governed regime (specimen I). This is consistent with the fact that the specimen is in axial equilibrium when material rate sensitivity dominates (specimen I) but the front face force $F_F$ and back face force $F_B$ are not equal once inertia dominates (specimen II).

\begin{figure}[htpb]
    \centering
    \includegraphics[width=1\textwidth]{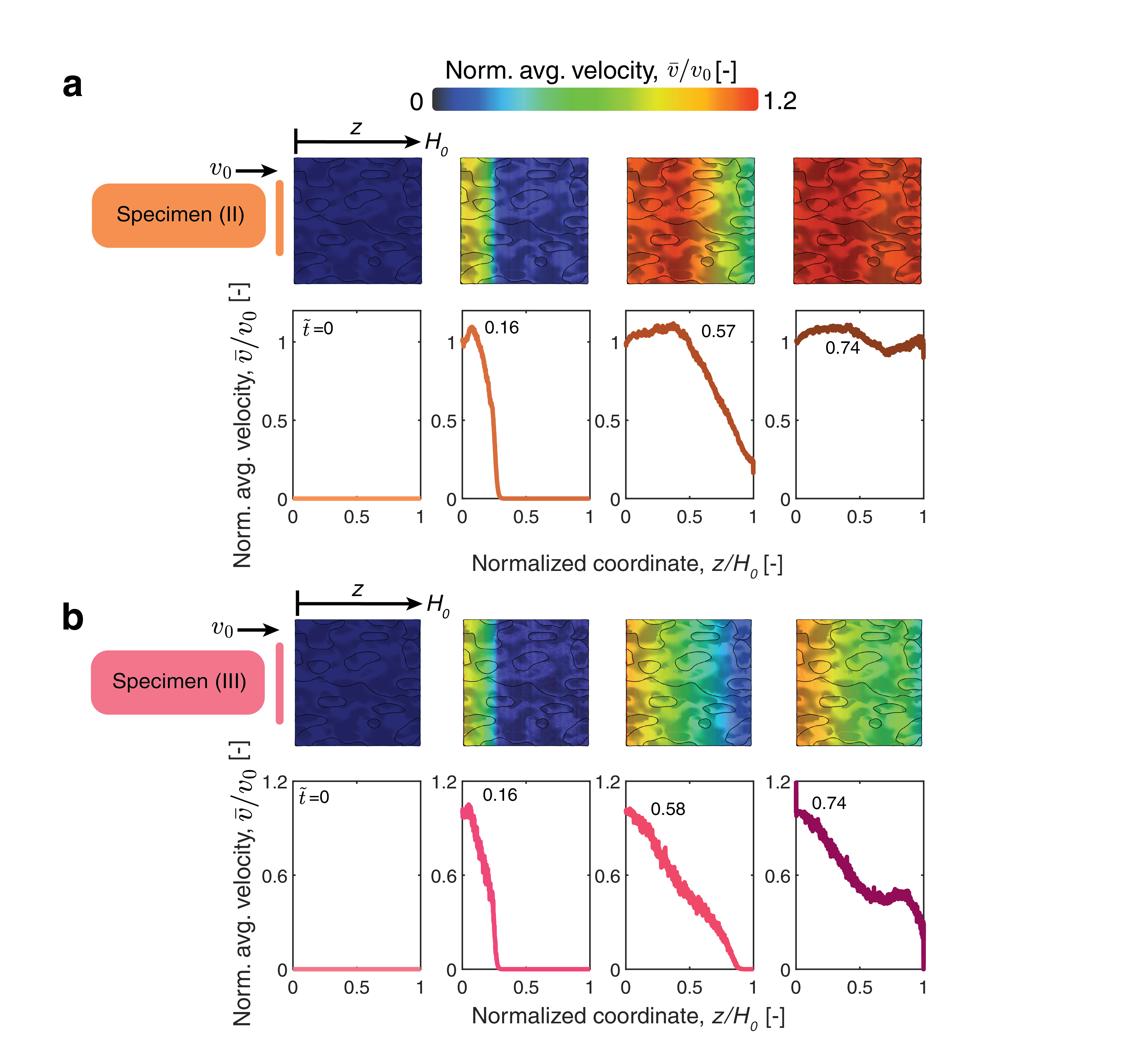}
    \caption{Predictions of the spatial distributions of the average velocity $\bar{v}$ at a given cross-sectional location $z$ in the undeformed configuration for (a) case (B) in Fig.~\ref{fig:dimensionlessmap}b and (b) case (C) in Fig.~\ref{fig:dimensionlessmap}b. The velocity $\bar{v}$ is normalized by the imposed deformation velocity $v_0$ and $z$ is defined in the inset of (a).}
    \label{fig:dispersion}
\end{figure}

Deep in the inertia-dominated regime (i.e., highest values of $\rho v_0^2/\sigma_0$) we would have anticipated the contours of $\sigma_B/\sigma_0$ to be nearly horizontal. However, we observe that the contours tend to become more vertical as $\rho v_0^2/\sigma_0$ is increased (Fig.~\ref{fig:dimensionlessmap}\hyperref[fig:dimensionlessmap]{a}), suggesting the existence of a third regime. To examine the behavior in this new regime, we consider a specimen (III) in the region marked (C) in Fig.~\ref{fig:dimensionlessmap}\hyperref[fig:dimensionlessmap]{b}. The differences in the mechanism of deformation  in the inertia-dominated regime (specimen II) and specimen (III) in region (C) can be illustrated by examining the velocity distributions within the two specimens. Let $z$ denote the axial location of a cross-section of the specimen in the undeformed configuration with $z=0$ corresponding to the displaced end of the specimen as illustrated in the first image of Fig.~\ref{fig:dispersion}\hyperref[fig:dispersion]{a}. We define $\bar{v}(z)$ as the average of the magnitude of the material-point velocity at a given cross-section located at $z$. Predictions of the distribution of the normalized velocities $\bar{v}/v_0$ as a function of $z/H_0$ are included in Figs.~\ref{fig:dispersion}\hyperref[fig:dispersion]{a,b} for specimens in regions (B) and (C), respectively. These distributions are shown for selected values of normalized time, $\tilde{t} \equiv c_0 t/H_0$, with elastic wave speed,  $c_0 = \sqrt{E^*/\rho}$, and the effective stiffness of the spinodal architecture assumed to be $E^* = E\cdot \bar{\rho}^2$. We observe that early in the deformation, the distribution of velocities in both specimens are similar, with the material primarily moving near the displaced end. However, as deformation progresses, we observe clear differences. While the velocity remains constant at  $\bar{v}\approx v_0$ near the displaced end in the inertia-dominated case (specimen II), the velocity distributions become more dispersed in specimen (III) with the velocity varying nearly linearly with $z$ throughout the specimen. This suggests strong dispersion effects are at play. Recall that the spinodal architected material specimens are heterogeneous with a complex topology comprising curved shells and internal boundaries, and thus we would expect the speeds of different elastic/plastic wave wavelengths to be differently affected, leading to dispersion. We therefore denote this third regime as a regime that is inertia-dominated with dispersion effects. This regime is marked in Fig.~\ref{fig:dimensionlessmap}\hyperref[fig:dimensionlessmap]{b} based on the demarcation where the contours of $\sigma_B/\sigma_0$ again tend to become more vertical at high values of $\rho v_0^2/\sigma_0$. This regime will be strongly influenced by the architected specimen topology and a detailed investigation of this regime is beyond the scope of the current study.

\subsection{Analogy to fluids}
In fluids, an important non-dimensional group is the Reynolds number $\mathcal{R}e \equiv \rho v_0 H_0/\eta$ which is the ratio of inertial to viscous forces, with $\eta$ denoting the fluid viscosity. This dimensionless group governs the transition from laminar flow, which is primarily viscosity governed, to turbulent flow, which is strongly influenced by inertia. We note that $\mathcal{R}e$ varies with a structural length scale, that we here are labeling $H_0$, leading to small $\mathcal{R}e$ for small length scales where flow is expected to be laminar and viscosity-governed. We have observed a similar phenomenon for our architected material specimens, with the response of the microscale specimens being governed by an effective viscosity---or rather strain-rate sensitivity of the constituent material---with inertia playing an important role in the macroscale specimens. From this analogy to fluids, we aim to determine whether a parameter akin to $\mathcal{R}e$ to also exists to describe this transition in the architected material specimens.

We define a non-dimensional group $\mathcal{M}\equiv \rho v_0 H_0/(\sigma_0/\dot{\varepsilon}_0)$. This is the ratio of the two groups that form the map in Figs.~\ref{fig:dimensionlessmap}\hyperref[fig:dimensionlessmap]{a,b}, viz. $\rho v_0^2/\sigma_0$ and $v_0/(H_0 \dot{\varepsilon}_0)$ are the inertial and viscous forces in our case. Interpreting $\sigma_0/\dot{\varepsilon}_0$ as the material viscosity, $\eta$, establishes the analogy between $\mathcal{M}$ and $\mathcal{R}e$. Contours of $\mathcal{M}$ are included in Fig.~\ref{fig:dimensionlessmap}\hyperref[fig:dimensionlessmap]{a}, which are necessarily straight and parallel lines as $\mathcal{M}$ is the ratio of the ordinate to abscissa of the map. Unlike $\mathcal{R}e$, which demarcates the transition from laminar to turbulent flow in linear viscous fluids, $\mathcal{M}$ is insufficient to demarcate the transition from constituent-material strain-rate sensitivity governed to inertia-governed responses for the spinodal architected specimens. This can be observed in Fig.~\ref{fig:dimensionlessmap}\hyperref[fig:dimensionlessmap]{a}, where moving along a constant contour of $\mathcal{M}$ traverses multiple regimes. The primary reason for this is that unlike a fluid, at vanishing strain rates, the architected specimen has an intrinsic strength approaching $\sigma_0$, which competes with the inertial forces $\rho v_0^2$. Consequently, instead of the single parameter, $\mathcal{M}$, the two groups $\rho v_0^2/\sigma_0$ and $v_0/(H_0 \dot{\varepsilon}_0)$ must be considered independently, forming a dimensionless map that describes the non-linear transition from material-rate sensitivity governed to inertia-governed responses of the architected specimens.

\subsection{Specimen length scale dependence}
Intrinsic material strength is not the only difference between the dynamics of solids and fluids, which can be made clearer by rearranging $\mathcal{M}$. In fluids, the Reynolds number $\mathcal{R}e$ implies the existence of a length scale $\eta/(\rho v_0)$ associated with the loading. Measurements suggest that flow is turbulent, with inertia playing an important role when $\mathcal{R}e$ exceeds a transition value $(\mathcal{R}e)_T$.  Depending on the topology of the structure,  $(\mathcal{R}e)_T$ takes values in the range $2000-10^4$. This transition value of $\mathcal{R}e$ implies a transition structural length scale $H_0 >(\mathcal{R}e)_T \eta/(\rho v_0)$ when inertial effects are important. In solids, there typically exists an intrinsic material length scale under dynamic conditions \cite{Needleman_1988} related to the elastic or plastic wave speeds. For example, in our spinodal architected materials the plastic wave speed, $c_p=\sqrt{\sigma_0/\rho}$, of the constituent material is set by the intrinsic material strength $\sigma_0$. This, combined with the strain-rate sensitivity sets the intrinsic material length scale $l \equiv (1/\dot{\varepsilon}_0)\sqrt{\sigma_0/\rho})$ such that $\mathcal{M}$ can be rewritten as  $\mathcal{M} = H_0 v_0/(l \,c_p)$. This length scale is independent of the loading rate $v_0$ and competes with the structural length scale $H_0$. Thus, the transition from constituent-material strain-rate sensitivity governed behavior to inertia-dominated behavior is dependent on the relations between $H_0/l$ and $v_0/c_p$, independently. 

For the material parameters in Table \ref{tab:MapProps}, $l\approx1.75\times10^5$  m and we include in Fig.~\ref{fig:dimensionlessmap}\hyperref[fig:dimensionlessmap]{b} contours of $H_0$ for this choice of $l$. Consider the $H_0=30$ mm and $100$ \textmu{}m contours which correspond to the two specimen sizes we have investigated in this study. The extent of these two contours spans the velocity range $1 \textnormal{ mm} \cdot\textnormal{s}^{-1}\leq v_0\leq 50 \textnormal{ m}\cdot \textnormal{s}^{-1}$. Over this velocity range we observe that the microscale specimens remain in the constituent-material strain-rate sensitivity governed regime while the macroscale specimens enter the inertia-dominated regime at high velocities. Thus, while $l \gg H_0$ has little significance in setting the regime of response of these specimens, the qualitative conclusion remains akin to fluids, viz. significantly higher velocities are required for the microscale specimens to become inertia dominated.

\section{Concluding remarks} \label{sec:Conclusion}
Scaling experiments to investigate relevant structural behavior have been a fundamental aspect of laboratory-scale fluid mechanics investigations. For example, the Reynolds number is typically used to characterize the transition from viscosity-dominated laminar flow to turbulent flow that is strongly influenced by inertia. Turbulence at large Reynolds numbers can occur either due to rapid flows or due to the large length scales of the structures. Using appropriate scaling laws, laboratory experiments are designed to investigate behaviors in the regimes of practical interest. Here, we attempt to develop scaling laws for architected material specimens using the spinodal architecture as an exemplar. The study employed a combination of experiments across a range of lengths and timescales, combined with extensive finite element (FE) calculations.

Polymeric specimens with a spinodal architecture were printed at micrometer (microscale) and millimeter (macroscale) length scales and tested over a wide range of imposed nominal strain rates
$\dot{\varepsilon}_n$. The measurements and FE calculations for the macroscale specimens show a dramatic increase in the peak strength with increasing strain rate for $\dot{\varepsilon}_n\gtrsim10^2$ s$^{-1}$. No such switch in behavior was observed for the microscale specimens. Detailed FE investigations revealed that this switch in behavior is associated with a transition from constituent-material strain-rate sensitivity governed response to inertia-dominated behavior in the macroscale spinodal specimens. This is akin to fluids, where turbulence, strongly influenced by inertia, occurs at lower flow velocities in large-scale structures.

For specimens made from a constituent material with a strength $\sigma_0$ at zero imposed strain rate, two non-dimensional groups describe the imposed loading of the specimens by a compressive velocity $v_0$. For specimens with an effective density $\rho$ and initial height $H_0$ these groups are: (i) the ratio of the inertial stress to the material strength given by $\rho v_0^2/\sigma_0$ and (ii) $v_0/(H_0 \dot{\varepsilon}_0)$, where $\dot{\varepsilon}_0$ is the reference strain rate that sets the constituent material strain-rate sensitivity and thus this group is the ratio of the imposed strain rate to a parameter that describes the constituent material strain-rate sensitivity. Unlike in fluids which do not possess an intrinsic strength $\sigma_0$ or intrinsic material length scale $l$, a single non-dimensional group does not describe the transition from constituent-material strain-rate sensitivity governed to inertia governed behavior.  Extensive FE calculations were used to map the regimes of behavior, and the results were presented as maps with these non-dimensional groups as axes. The maps clearly demonstrate that, for a given range of imposed velocities, the macroscale specimens are more likely to transition to inertia-governed behavior, whereas the microscale specimens are expected to remain dominated by constituent material strain-rate sensitivity.

Our experimental and numerical investigation used the spinodal architecture as an exemplar. While the quantitative results are valid for this topology and the polymeric constituent material used in our study, the qualitative results are expected to be more general. In particular, the two non-dimensional groups $\rho v_0^2/\sigma_0$ and $v_0/(H_0 \dot{\varepsilon}_0)$ will set the regime of response of the specimens and these groups can be used to design laboratory scale specimens whose response lies in the regime of structural interest.

\section*{Acknowledgments}
V.S.D. acknowledges funding from the UKRI Frontier Research grant ``Graph-based Learning and design of Advanced Mechanical Metamaterials'' with award number EP/X02394X/1 and the Office of Naval Research project ``Adaptable mechanical metamaterials with tailorable toughness and energy absorption'' with award number N00014-23-1-2797. C.M.P. acknowledges partial financial support from DEVCOM Army Research Laboratory Army Research Office through the Massachusetts Institute of Technology (MIT) Institute for Soldier Nanotechnologies (ISN) under Cooperative Agreement number W911NF-23-2-0121, and from the National Science Foundation (NSF) CAREER Award (CMMI-2142460). This work was carried out in part through the use of MIT.nano facilities.

\vspace{-1em}
	\bibliographystyle{elsarticle-num}
	{\footnotesize\bibliography{References}}
\end{document}